\documentclass[twocolumn,pre,aps]{revtex4}

\usepackage{graphicx}
\usepackage{dcolumn}
\usepackage{bm}

\begin{document}

\title{Reconstruction of Plane Circular Currents
from Their Orthogonal Magnetic Field}

\author{Yu.\,E.\,Kuzovlev}
\email{kuzovlev@kinetic.ac.donetsk.ua} \affiliation{Donetsk
Physics and Technology Institute, 83114 Donetsk, Ukraine}

\date{\today}

\begin{abstract}
The solution is presented to the problem what distribution of
electric current in thin circular film provides a given
distribution of normal (perpendicular to film) component of the
current-induced magnetic field at film's surface.
\end{abstract}


\maketitle

\section{Introduction}
Plane electric current distributions and magnetic fields produced
by them are of great interest in a number of applications, for
instance, when investigating magnetic flux pinning and critical
states in thin superconductor films. At given current, one can
calculate its field anywhere by direct integration. But in
practice the inverse problem arises: how one could extract the
current from measuring orthogonal (normal to plane) component of
its field in some film's neighborhood, e.g. at its surface?
Although this is standard problem of the classical theory of
potential, I did not find its solution in textbooks. Therefore I
ventured to derive it independently, for special case of axial
symmetry.

\section{The problem}
Let a film with radius $\,R\,$ and small thickness $\,D<<R\,$
($D/R\rightarrow 0$) lies in the $XY$-plane ($\,\rho\equiv
\sqrt{x^2+y^2}<R\,$, $\,|z|<D/2\,$) and carries a circular current
whose density integrated over the thickness is $\,J(\rho)\,$. We
assume that in fact we know not $\,J(\rho)\,$ itself but normal
component of its magnetic field close to the film's surface,
$\,H_z(\rho,z=\pm D/2\rightarrow 0)\,$. In case of superconductor
film, this quantity represents density of vortices which pierce
the film. From known $\,H_z(\rho,0)\,$ we want to obtain
tangential radial component of same field,
$\,H_{\rho}(\rho,z=D/2\rightarrow +0)=$
$-H_{\rho}(\rho,z=-D/2\rightarrow -0)\,$. If the latter is known,
then the integral surface current density $\,J(\rho)=$ $\int
j_{\phi}(\rho,z)dz\,$ (with $\,j_{\phi}\,$ being azimuth current
density) can be found with the help of relation
\[
\frac {4\pi}{c}J(\rho)=\pm\, 2H_{\rho}(\rho,\pm D/2)-\frac
{\partial}{\partial\rho}\int_{-D/2}^{D/2}H_z(\rho,z)dz\approx
\]
\begin{equation}
\approx\, \pm\, 2H_{\rho}(\rho,\pm 0)\,\,\,,\label{J}
\end{equation}
which follows from the Maxwell equation ~{\bf
curl}$\,\bm{H}=4\pi\bm{j}/c\,\,$ (written in the CGSE units).

Outside the film, the field has potential character,
$\,\bm{H}=-\nabla U\,$, with a potential $\,U\,$ satisfying the
Laplace equation $\,\triangle U=0\,$. With reference to circular
geometry, it is natural to consider the potential $\,U\,$ using
the spheroidal coordinates. Their definition and examples of there
application can be found e.g. in \cite{ll}. We use slightly
modified version of spheroidal coordinates, in the form
\begin{equation}
\begin{array}{c}
\rho^2=R^2(1+u^2)(1-v^2)\,\,,\,\,\,z^2=R^2u^2v^2\,\\
\,\,\,(\,0<u^2<\infty\,\,,\,\,\,0<v^2<1\,)\,\,, \label{sc}
\end{array}
\end{equation}
at that the horizontal angle stays the third coordinate:
$\,x=\rho\cos\,\phi\,$, $\,y=\rho\sin\,\phi\,$. To write the
Laplace operator, $\,\triangle \,$, in these coordinates, it is
convenient to introduce two operators
\begin{eqnarray}
\Lambda_{-}(u)\equiv \frac {\partial}{\partial u}(1+u^2)\frac
{\partial}{\partial u}\,\,,\,\,\,\Lambda_{+}(v)\equiv \frac
{\partial}{\partial v} (1-v^2)\frac {\partial}{\partial
v}\,\label{lo}
\end{eqnarray}
Then $\,\triangle \,$ looks as
\begin{equation}
\triangle=-\frac
{1}{R^2\,(u^2+v^2)}\,\left[\Lambda_{-}(u)+\Lambda_{+}(v)\right]+\,\frac
{1}{\rho^2}\,\frac {\partial^2 }{\partial \phi^2}\,\,,
\label{laplace}
\end{equation}
The film occupies the region $\,\rho<R$, $z=0\,$, which
corresponds to $\,u=0\,$, while the rest of the plane $\,z=0\,$ to
$\,v=0\,$. Of course, the potential must be anti-symmetric with
respect to this plane, therefore,
\begin{equation}
U(\rho>R,z=0)=U(u,v=0)=0  \label{zero}
\end{equation}
At film's surface the potential has discontinuity, but its normal
derivative is continuous. By the terms of our task,
\[
\left[-\frac {\partial}{\partial z}\,U(\rho<R,z)\right]_{z=0}
=\left[-\frac {1}{Rv}\,\frac {\partial}{\partial
u}\,U(u,v)\right]_{u=0}=
\]
\begin{equation}
=H_{z0}(\rho)=H_{z0}\left(R\sqrt{1-v^2}\right)\,\,, \label{bc}
\end{equation}
where $\,H_{z0}(\rho)=H_{z}(\rho<R,0)\,$ is a given function.
Formally, the problem $\,\triangle U=0\,$ with boundary conditions
(\ref{zero}) and (\ref{bc}) is equivalent to the problem about
electric potential of a charge distributed, with density
$\,H_{z0}\delta(z)/2\pi\,$, in the circular hole $\,\rho<R\,$ cut
out of ideally conducting plane $\,z=0$.

If the axial symmetry is assumed, then the Laplace equation
reduces to
\begin{eqnarray}
(\Lambda_{-}+\Lambda_{+})\,U=\,0\,\,\label{le}
\end{eqnarray}

\section{Magnetic potential}
The operator $\,\Lambda_{+}\,$ in (\ref{le}) is nothing but
generating operator for the complete set of orthogonal (at
interval $\,-1<v<1\,$) Legendre polynomials $\,P_n(v)\,$ (see e.g.
\cite{nu}):
\begin{eqnarray}
\Lambda_{+}(v)P_n(v)=-\lambda_n P_n(v)\,\,, \,\,\,\lambda_n =
n(n+1)\,\,,\label{lp}
\end{eqnarray}
Therefore, particular solutions to Eq.\ref{le} have the form
$\,P_n(v)Q_n(u)\,$, where functions $\,Q_n(u)\,$ satisfy the
equation $\,\Lambda_{-}(u) Q_n(u)=\lambda_n Q_n(u)\,$. It is easy
to see that $\,\Lambda_{-}(iv)=$ $-\Lambda_{+}(v)\,$.
Consequently, $\,Q_n(u)\propto \widetilde{P}_n(iu)\,$, with
$\,\widetilde{P}_n(u)\,$ also being solutions of Eq.\ref{lp}:
$\,\Lambda_{+}(u) \widetilde{P}_n(u)=-\lambda_n
\widetilde{P}_n(u)\,$. That must be chosen be the second kind
(non-polynomial) eigenfunctions which have zero asymptotic at
infinity, because $\,U(u,v)\,$ should tend to zero at
$\,z\rightarrow\infty$. Hence, according to general theory of
classical special functions \cite{nu}, $\,Q_n(u)\,$ can be
represented by
\begin{eqnarray}
Q_{2k}=\int_{-1}^{1}\frac {uP_{2k}(s)ds}{s^2+u^2}\,\,,\,\,\,
Q_{2k+1}=\int_{-1}^{1}\frac {sP_{2k+1}(s)ds}{s^2+u^2} \label{ss}
\end{eqnarray}

In view of the boundary condition (\ref{zero}), as well as
(\ref{bc}), the odd polynomials with $\,n=2k+1\,$ ($k=0,1,..$)
only contribute to complete solution of Eq.\ref{le}. By this
reason, we can write
\begin{equation}
U(u,v)=\sum_{k=0}^{\infty}\,(4k+3)\,C_k\,P_{2k+1}(v)\,
Q_{2k+1}(u)\,\,,\label{exp}
\end{equation}
with coefficients $\,C_k\,$ to be determined from condition
(\ref{bc}). Notice \cite{nu} that $\,P_{2k+1}(v)\,$ form complete
orthogonal set in the half-interval $\,0<v<1\,$. Under standard
classical definition of Legendre polynomials,
\begin{equation}
\begin{array}{c}
\int_0^1 P_{2k+1}(s)P_{2m+1}(s)\,ds=\delta_{km}/(4k+3)\,\,,
\label{norm}
\end{array}
\end{equation}
that is $\,4k+3\,$ in (\ref{exp}) serves as normalizing
multiplier.

The result of subsequent proper calculations (detailed in Appendix
in Sec.1) reads
\begin{equation}
C_k=(-1)^kS_k\int_0^R P_{2k+1}\left (\sqrt{1-\frac
{r^2}{R^2}}\right )H_{z0}(r)\frac {r\,dr}{\pi R} \label{coefs}
\end{equation}
Here the quantities
\begin{equation}
\begin{array}{c}
S_k\,\equiv\, B(1/2,k+1)/2\,=\,2^k k!/(2k+1)!!\,\,\label{sk}
\end{array}
\end{equation}
are introduced. At upper side of film's surface, formulas
(\ref{exp}) and (\ref{coefs}), as combined with (\ref{ss}) and
(\ref{sc}), yield (see Appendix 2)
\begin{equation}
U(\rho,z=+0)=\int_0^R G_R(\rho,r)\,H_{z0}(r)\,r\,dr\,\,,
\label{sp}
\end{equation}
where the Green function $\,G_R(\rho,r)\,$ is presented by
\begin{widetext}
\begin{equation}
G_R(\rho,r)\,\equiv \,\frac {2}{\pi
R}\,\sum_{k=0}^{\infty}\,(4k+3)\,S_k^2\, P_{2k+1}\left
(\sqrt{1-\frac {\rho^2}{R^2}}\right )P_{2k+1}\left (\sqrt{1-\frac
{r^2}{R^2}}\right )\,= \label{series}
\end{equation}
\begin{equation}
=\,\frac {2}{\pi\max(\rho,r)}\left[\,K\left(\frac
{\min(\rho,r)}{\max(\rho,r)}\right
)-F\left(\arcsin\frac{\max(\rho,r)}{R}\,,\,\,\frac
{\min(\rho,r)}{\max(\rho,r)}\,\right )\,\right]\,= \label{sum}
\end{equation}
\begin{equation}
=\,\frac {2}{\pi\,(\rho +r)}\left[\,K\left(\frac {2\sqrt{\rho
r}}{\rho +r}\right )-2F\left(\frac 12 \arcsin\frac {\rho}{R}+\frac
12 \arcsin\frac {r}{R}\,,\,\,\frac {2\sqrt{\rho r}}{\rho
+r}\,\right )\,\right]\,\,\,, \label{tr}
\end{equation}
\end{widetext}
where $\,F(\phi,k)\,$ is the first-kind elliptic integral and
$\,K(k)=F(\pi/2,k)\,$ the first-kind complete elliptic integral.
The way of summation of the series (\ref{series}) is described in
Sec.3 of Appendix.

\section{Current}
Provided the normal field at surface of (infinitely) thin film is
known, the current can be restored by means of (\ref{J}) and
(\ref{sp}):
\begin{equation}
\frac {2\pi}{c}J(\rho)=H_{\rho 0}(\rho)=-\frac {d}{d\rho}\int_0^R
G_R(\rho,r)\,H_{z0}(r)\,r\,dr\,\,, \label{curr}
\end{equation}
where $\,H_{\rho 0}(\rho)\equiv H_{\rho}(\rho,z=+0)\,$.
Integration of (\ref{curr}) yields the mean current:
\begin{equation}
\frac {2\pi}{c}\,\frac {1}{R}\int_0^RJ\,d\rho =\frac
{1}{R}\int_0^R\left(1-\frac {2}{\pi}\arcsin\frac {r}{R}\right
)\,H_{z0}(r)\,dr\,\, \label{mc}
\end{equation}
It is easy to calculate also the magnetic moment, $\,M\,$, of the
current~\cite{mk}:
\begin{equation}
\begin{array}{c}
M=\frac {2}{\pi}\int_0^Rr\,\sqrt{R^2-r^2}\,H_{z0}(r)\,dr
\label{mm} \
\end{array}
\end{equation}
Thus both these characteristics eliminate contributions from the
film's edge, i.e. $H_{z0}(\rho \approx R)\,$.

The expansion (\ref{sum}), when substituted to (\ref{curr}),
determines ``natural modes'' of the surface field:
\begin{eqnarray}
H_{z0}=h_{\perp k}(\rho )\equiv \frac
{P_{2k+1}(\sqrt{1-(\rho/R)^2})}{\sqrt{1-(\rho/R)^2}}\,\,
\Leftrightarrow\,\,\label{nm1} \\ \Leftrightarrow\,H_{\rho
0}=h_{\parallel k}(\rho)\equiv \frac {2}{\pi}\,S_k^2\,\frac
{\,\rho\, P^{\prime}_{2k+1}(\sqrt{1-(\rho/R)^2})}
{\sqrt{R^2-\rho^2}}\, \label{nm2}
\end{eqnarray}
Clearly, $\,h_{\perp k}(\rho )\,$ is $\,2k$-order polynomial, and
$\,h_{\perp k}(0)=1\,$ (because $\,P_n(1)=1$), while
$\,h_{\parallel k}(\rho )\,$ is polynomial of the same order
multiplied by factor $\,\rho/\sqrt{R^2-\rho^2}\,$ and possesses
square-root divergency at film's edge. Both $\,h_{\perp k}(\rho
)\,$ and $\,h_{\parallel k}(\rho )\,$ have $\,k\,$ zeros at
$\,0<\rho<R\,$. In particular, $\,h_{\parallel 0}(\rho )=$
$2\rho/\pi\sqrt{R^2-\rho^2}\,$ determines the current distribution
(see e.g. \cite{mk}) which creates constant unit-value field at
film's surface. Of course, one can form linear combinations of
these modes without current's divergency at the edge. For example,
the two lowest modes ($\,k=0,1\,$) give
\begin{equation}
\begin{array}{c}
H_{z0}=1-\frac {3}{2}\,x^2\,\,\Leftrightarrow\,\,H_{\rho 0}=\frac
{4}{\pi}\,x\,\sqrt{1-x^2}\,\,\,, \label{nd}
\end{array}
\end{equation}
with $\,x\equiv \rho/R\,$.

The approach mentioned in Appendix 3 allows to divide (\ref{curr})
into singular and regular parts:
\begin{equation}
H_{\rho 0}(\rho)\,=\,H_{\rho 0}^{sing}(\rho)\,+\,H_{\rho
0}^{reg}(\rho)\,\,,\label{sre}
\end{equation}
\begin{equation}
H_{\rho 0}^{sing}(\rho)=\frac
{2\,\rho}{\pi\sqrt{R^2-\rho^2}}\int_0^R\frac
{H_{z0}(r)\,r\,dr}{R\sqrt{R^2-r^2}}\,\,,\label{sing}
\end{equation}
\begin{equation}
H_{\rho 0}^{reg}(\rho)=-\,\int_0^R T\left(\frac {\rho}{R},\frac rR
\right )\frac rR\,\,\frac {dH_{z0}(r)}{dr}\,\,dr\,\,\label{curr1}
\end{equation}
In the latter formula,
\begin{equation}
T(x,y)\equiv \frac
{2\,[K(k)-F(\phi,k)-E(k)+E(\phi,k)]}{\pi\min(x,y)}\,\,,\label{tf}
\end{equation}
with $\,E(k)\,$ and $\,E(\phi,k)\,$ being the second-kind elliptic
integrals whose module of ellipticity $\,k\,$ and phase $\,\phi\,$
are expressed by
\begin{equation}
\begin{array}{c}
k\equiv \min(x,y)/\max(x,y)\,\,,\,\,\,\phi\equiv \arcsin\max(x,y)
\label{tf2}
\end{array}
\end{equation}

Clearly, $\,T(x,y)\,$ has logarithmic peak at $\,x=y\,$ (see
Fig.1). What is important, $\,T(1,y)=T(x,1)=0\,$, therefore
$\,H_{\rho 0}^{reg}(\rho \rightarrow R)\,$ is zero (or at least
finite value). In opposite, $\,H_{\rho 0}^{sing}(\rho \rightarrow
R)\,$ diverges except the cases when the integral in (\ref{sing})
turns into zero (e.g. in the example (\ref{nd})). In such case,
(\ref{curr}) becomes integral relation (\ref{curr1}) between
current and radial derivative of normal field (in place of
differential relation $\,4\pi j/c=-\partial H_z/\partial\rho\,$
which would work under cylindrical geometry).
\begin{figure}
\includegraphics{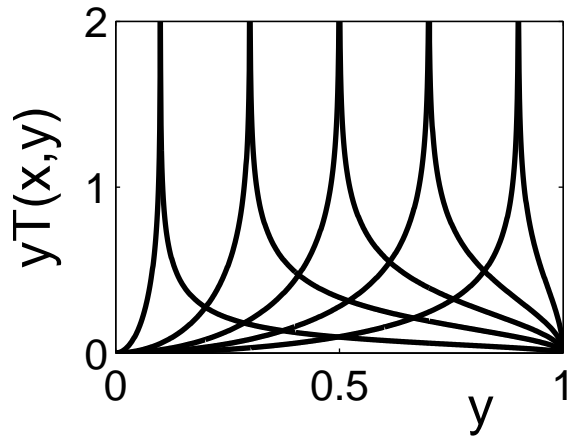}
\caption{\label{fig1} The kernel $\,T(x,y)\,y\,$ of the integral
relating the current to the field gradient, as function of $\,y\,$
at $\,x\,=\,0.1,\,0.3,\,0.5,\,0.7$ and $0.9\,$. }
\end{figure}

\section{Conclusion}
To resume, we found how to determine an electric current
distribution in thin circular film from normal component of the
current-induced field measured in the vicinity of film's surface
(very similar task about infinite strip film must be considered
separately). The results can be applied, in particular, to numeric
modeling of magnetic flux penetration into superconducting film.

I am grateful to Dr.~Yu.\,Medvedev and Dr.~V.\,Khokhlov for useful
comments.

\appendix*
\section{}

{\bf 1}. According to (\ref{bc}), (\ref{exp}) and (\ref{norm}),
\begin{equation}
\begin{array}{c}
C_k=-[Q_{2k+1}^{\prime}(+0)]^{-1}\,\times \label{a11}\\
\times \,R\,\int_0^1
s\,P_{\,2k+1}(s)\,H_{z0}\left(R\sqrt{1-s^2}\right)ds\,,
\end{array}
\end{equation}
where the upper half-space $\,z>0\,$ is considered, and the prime
means differentiation:
\begin{equation}
\begin{array}{c}
Q_{2k+1}^{\prime}(+0)\,\equiv\, \lim_{\,u\rightarrow +0}\,\,d
Q_{2k+1}(u)/du\, \label{a12}
\end{array}
\end{equation}
From (\ref{ss}) it follows that at
\[
Q_{2k+1}(u)=Q_{2k+1}(0)-\int_{-1}^1\frac
{u^2P_{2k+1}(s)ds}{s(u^2+s^2)} \label{a13}
\]
Here in the limit $\,u\rightarrow+0\,$ factor $\,u/(u^2+s^2)\,$
works as $\,\pi\delta(s)\,$, therefore
\begin{equation}
Q_{2k+1}^{\prime}(+0)=-\pi\lim_{s\rightarrow
0}P_{2k+1}(s)/s=-\pi\,P_{2k+1}^{\prime}(0) \label{a13}
\end{equation}
The right-hand sides of these equalities can be easily found from
the generating function of Legendre polynomials \cite{nu},
\begin{equation}
\sum_{n=0}^{\infty} t^nP_n(v)=1/\sqrt{1+t^2-2tv}\,\, \label{gf}
\end{equation}
Its differentiation by $\,v\,$ at point $\,v=0\,$ and Taylor
expansion over $\,t\,$ show that
\begin{equation}
\begin{array}{c}
P_{2k+1}^{\prime}(0)\,=\,(-1)^k/S_k\,\,\,, \label{a14}
\end{array}
\end{equation}
with $\,S_k\,$ defined by (\ref{sk}), and
$\,P_{2k+1}^{\prime}(s)\equiv $ $dP_{2k+1}(s)/ds\,$. Combining
(\ref{a11}), (\ref{a13}) and (\ref{a14}) and performing the
obvious change of variables in the integral in (\ref{a11}), we
obtain (\ref{coefs}).

{\bf 2}. According to (\ref{exp}) and (\ref{sc}), to express the
potential at upper film's side, $\,U(\rho,z\rightarrow +0)=$
$U(u\rightarrow +0,v=\sqrt{1-\rho^2/R^2})\,$, we should calculate
the quantities $\,Q_{2k+1}(0)\,$ determined by (\ref{ss}). This
can be made either with the help of the recurrent relations
between Legendre polynomials \cite{nu} or with the help of
(\ref{gf}). Direct integration of (\ref{gf}), after dividing by
$\,v\,$ and replacing $\,t\,$ by $\,it\,$, yields
\begin{equation}
\frac {1}{2i}\,\sum_{k=0}^{\infty}\,(it)^{2k+1}\int_{-1}^1\frac
{P_{2k+1}(v)}{v}\,dv=\frac {\arcsin\,t}{\sqrt{1-t^2}}\,
\label{sgf}
\end{equation}
From another hand, let us pay attention to that the coefficients
(\ref{sk}) can be represented as
\begin{equation}
\begin{array}{c}
S_k=\int_0^{\pi/2} \sin^{2k+1}\theta\, d\theta\,\,\,, \label{a22}
\end{array}
\end{equation}
therefore their generating function coincides with (\ref{sgf}):
\begin{equation}
\sum_{k=0}^{\infty}\,t^{2k+1}S_k=\int_0^{\pi/2}\frac
{t\sin\,\theta\,d\theta }{1-t^2\sin^2\theta}=\frac
{\arcsin\,t}{\sqrt{1-t^2}}\, \label{skgf}
\end{equation}
Comparing  (\ref{skgf}) and (\ref{sgf}), and taking into account
(\ref{ss}) and (\ref{sk}), we conclude that
\begin{equation}
Q_{2k+1}(0)=\int_{-1}^1\frac {P_{2k+1}(s)}{s}\,ds\,=\,2(-1)^kS_k
\label{a21}
\end{equation}

{\bf 3}. Let us consider the series
\begin{equation}
\sigma (x,y)\,\equiv\,\sum_{k=0}^{\infty}\,(4k+3)\,S_k^2\,P_{2k+1}
(x)P_{2k+1}(y)\,\,, \label{ser}
\end{equation}
which turns into (\ref{series}) after the evident change of
variables (and adding constant multiplier). One of ways to sum it
is based on the remarkable formula \cite{vi}:
\begin{equation}
\begin{array}{c}
\sum_{n=0}^{\infty}\,(2n+1)\,P_n(x)P_n(y)P_n(z)\,=\\
\,\label{vil} \\=\,\frac {2}{\pi}\,\,\mathbf{Re}\,\,
1/\sqrt{1-x^2-y^2-z^2+2xyz}\,\,\,,
\end{array}
\end{equation}
where $\,-1<x,y,z<1\,$ and the symbol $\,\mathbf{Re}\,$ means
taking real part, that is the right-hand side is zero when the
subradical expression is negative (in \cite{vi} this formula is
presented in slightly different notations).

Now, divide both sides of (\ref{vil}) by factor $\,2z\,$ and
integrate over $\,z\,$ from $\,-1\,$ to $\,1\,$ (in the sense of
principal value). On the right-hand side we have exactly
calculable integral. On the left, apply the relation (\ref{a21})
(clearly, the odd terms only survive, with indices $\,n=2k+1$).
The result is
\begin{equation}
\begin{array}{c}
g(x,y)\,\equiv\,\sum_{n=0}^{\infty}\,(4k+3)(-1)^kS_k\,
P_{2k+1}(x)P_{2k+1}(y)\,=\\
\,\label{a31}
\\=\,\,\mathbf{Re}\,\,\,\mathrm{sign}(xy)\,/\sqrt{x^2+y^2-1}\,\,
\end{array}
\end{equation}

Next, multiply together $\,g(x,s)\,$ and $\,g(y,s)\,$ and
integrate the product over $\,0<s<1\,$. On the left we should
apply the orthogonality and normalization relations (\ref{norm}).
On the right, we arrive to a standard elliptic integral,
eventually obtaining the equality
\begin{equation}
\sigma (x,y)\,=\,\frac
{\mathrm{sign}(xy)\,[K(k)-F(\phi,k)]}{\sqrt{1-\min
(x^2,y^2)}}\,\,, \label{res}
\end{equation}
where  $\,K(k)\,$ and $\,F(\phi,k)\,$ are standardly designated
first-kind elliptic integrals (complete and incomplete,
respectively) whose module of ellipticity $\,k\,$ and phase
$\,\phi\,$ are expressed by
\[
\begin{array}{c}
\phi\, \equiv \,\arcsin\sqrt{1-\min (x^2,y^2)}\,\,\,,\\\,\\
k\,\equiv \,\sqrt{[1-\max (x^2,y^2)]\,/\,[1-\min
(x^2,y^2)]}\,\,\,,
\end{array}
\]
and function $\,\sigma (x,y)\,$ is defined by (\ref{ser}).
Changing here $\,x\,$ and $\,y\,$ by $\,\sqrt{1-\rho^2/R^2}\,$ and
$\,\sqrt{1- r^2/R^2}\,$, respectively, and multiplying the result
by $\,2/\pi R\,$, we come to the relations
(\ref{series})-(\ref{sum}). The conversion of (\ref{sum}) into
(\ref{tr}) is based on properties of elliptic integrals (see e.g.
\cite{kk,dw}).

\end{document}